\documentclass[a4paper,fleqn,usenatbib,useAMS]{mnras}

\usepackage{graphicx}	
\usepackage{amsmath}	
\usepackage{amssymb}	
\usepackage{multicol}        
\usepackage{bm}		
\usepackage{pdflscape}	
\usepackage{ulem}	%

\newcommand{\kms}{\,km\,s$^{-1}$} 

\usepackage[T1]{fontenc}
\usepackage{ae,aecompl}

\usepackage{txfonts}

 \def\target  {LS~I~+61$^\circ$303 }
 \def\mqso  {microquasar}

\title[Revisiting LS~I+61$^\circ$303 with VLBI astrometry]{\textit{Revisiting \target with VLBI astrometry}}

\author[Y. W. Wu]{\Large {Y. W. Wu$^{1,2}$\thanks{\href{mailto:yuanwei.wu@ntsc.ac.cn}{yuanwei.wu@ntsc.ac.cn}} , G. Torricelli-Ciamponi$^{3}$ M. Massi$^{4}$\thanks{\href{mailto:mmassi@mpifr-bonn.mpg.de}{mmassi@mpifr-bonn.mpg.de}},M. J. Reid$^{5}$, B. Zhang$^{6}$, L. Shao$^{7}$, X. W. Zheng$^{8}$}
\\
$^{1}$ National Astronomical Observatory of Japan, Osawa 2-21-1, Mitaka, Tokyo 181-8588, Japan\\
$^{2}$ National Time Service Center, Key Laboratory of Precise Positioning and Timing Technology, Chinese Academy of Sciences, Xi'an 710600, China\\
$^{3}$ INAF - Osservatorio Astrofisico di Arcetri, L.go E. Fermi 5, Firenze, Italy\\
$^{4}$ Max-Planck-Institut f\"{u}r Radioastronomie, Auf demH\"{u}gel 69, 53121 Bonn, Germany\\
$^{5}$ Harvard-Smithsonian Center for Astrophysics, 60 Garden Street, Cambridge, MA 02138, USA\\
$^{6}$ Shanghai Astronomical Observatory, 80 Nandan Road, Shanghai 20030, China\\
$^{7}$ Max Planck Institute for Gravitational Physics (Albert Einstein Institute),Am M\"{u}hlenberg 1, D-14476 Potsdam-Golm, Germany \\
$^{8}$ School of Astronomy and Space Sciences of Nanjing University, Nanjing 210093, China }
\date{Last updated 2017 April 15}

\pubyear{2017}

\begin{document}
\label{firstpage}
\pagerange{\pageref{firstpage}--\pageref{lastpage}}
\maketitle

\begin{abstract}
 We conducted multi-epoch VLBA phase reference observations of \target 
 in order to study its precessing radio jet.  Compared to similar observations
 in 2006, we find that the observed elliptical trajectory of emission at 8.4 GHz 
 repeats after the 9-year gap.  The accurate alignment of the emission patterns 
 yields a precession period of 26.926 $\pm$ 0.005 d, which is consistent with
 that determined by Lomb-Scargle analysis of the radio light curve. 
 We analytically model the projection 
on the sky plane of the peak position of a precessing, synchrotron-emitting jet,
which traces an elliptical trajectory on the sky.
Comparing the simulation with the VLBA astrometry we improve our knowledge 
of the geometry of the system.
 We measure
 the \target absolute proper motion to be $-0.150$~$\pm$~0.006~mas~yr$^{-1}$ eastward
 and $-$0.264~$\pm$~0.006~mas~yr$^{-1}$ northward.  Removing Galactic rotation,
 this reveals a small, $<20$~km~s$^{-1}$, non-circular motion, which indicates
 a very low kick velocity when the black hole was formed.  
\end{abstract}




\begingroup
\let\clearpage\relax
\tableofcontents
\endgroup
\newpage

 \section{Introduction}
\target is an unusual high-mass x-ray binary (HMXB). It was discovered as a
periodic radio source by \citet{1978Natur.272..704G} and
later periodic variations in X-ray \citep{1997AA...320L..25P,
2001ApJ...561.1027E} and $\gamma$-ray \citep{2009ApJ...693..303A} were
observed.  The origin of its $\gamma$-ray, X-ray, optical/infrared and radio
emission has been debated for decades
\citep{1992ApJ...395..268T, 1995A+A...298..151M, 2006A+A...459L..25B,
2007A+A...474...15R, 2013ASPC..470..373M}. The X-ray characteristics of
\target fit those of accreting black holes at moderate luminosity
\citep{2017arXiv170401335M} that would make this source along with MWC 656,
the only systems where a black hole accretes from the wind of
a Be companion star.
 
\citet{2002A+A...385L..10K} suggested  \target is a precessing microblazar, a
kind of microquasars where precession periodically brings the approaching jet
close to the line of sight and Doppler boosting its emission.  Very Long
Baseline Array (VLBA) astrometry in 2006 provided the first estimate of the
precession period.  In order to study the properties of the precessing jet, we
conducted a second set of multi-epoch VLBA phase-reference observations in
2015.  Section 2 summarizes previous radio  observations of \target and the
precessing scenario.  We describe the new observations and data reduction in
Section 3. In Section 4 we present a joint analysis of 2006 and 2015 VLBA
datasets, which indicates a very stable precession of \target and allows us to
determine an accurate precession period.  In Section 5 we present our
theoretical astrometry model.  In Section 6, we discuss the measured
3-dimensional motion of \target.  Section 7 presents our conclusions.
\section{The precessing jet in \target}

\begin{figure}
  \includegraphics[width=8cm, angle=0]{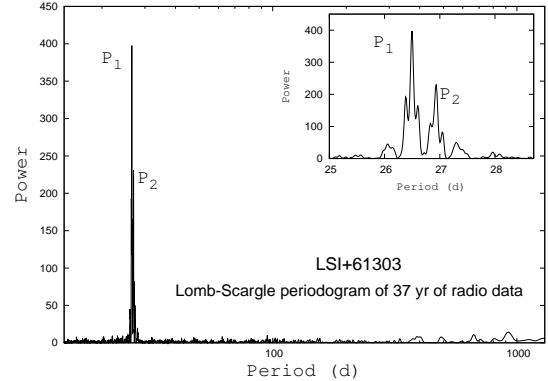}
 \caption{Lomb-Scargle periodogram of LSI+61303 with  zoom of the spectral window 25--29 d around $P_1$ and $P_2$.}
 \label{fig-LB}
 \end{figure}

 A Lomb-Scargle timing analysis of  36.8-yr  radio observations   \citep{2016A+A...585A.123M}
confirmed previous discoveries \citep[e.g.,][]{2015A+A...575L...9M} 
of two characteristic periods, $P_1~=~26.496~\pm~0.013$~d and $P_2~=~26.935~\pm~0.013$~d, in the emission from \target
(see Fig.  \ref{fig-LB}).
The period $P_1$ corresponds to orbital periodicity \citep{2002ApJ...575..427G}.  
Several authors \citep{1992ApJ...395..268T, 1995A+A...298..151M, 2006A+A...459L..25B,
2007A+A...474...15R, 2016A+A...595A..92J} 
have shown that because of the high eccentricity  ($e$~=~0.7, \citet{2005MNRAS.360.1105C}) 
there are two accretion peaks along the orbit of \target: 
one close to periastron and a second one shifted towards apastron.
Near periastron 
the ejected relativistic particles encounter the strong stellar radiation field of the B0 star
and suffer strong inverse-Compton (IC) losses and, thus, do not produce a radio outburst.
However, for the second accretion peak near apastron, the IC losses are smaller and 
synchrotron emission in the radio band is observed
\citep{1992ApJ...395..268T, 1995A+A...298..151M, 2006A+A...459L..25B,
2007A+A...474...15R, 2016A+A...595A..92J}. 

The  second feature in the  spectrum of Fig. 1,
at $P_2~=~26.935~\pm~0.013$~d,  was more challenging to understand.  The simplest explanation
is that the observed flux density from a relativistic jet \citep{1999ARA+A..37..409M} 
is the product of an intrinsically variable jet and Doppler boosting toward the observer: 
$S_{\rm {observed}}~=~S_{\rm {intrinsic}}(f(P_1)) \times DB$, where $DB$ is the Doppler
boosting factor.  \citet{2014A+A...564A..23M} suggested that the $DB$ factor could be a 
function of $P_2$.

The observations indeed support a variation of the jet angle. European VLBI Network (EVN),
MERLIN and VLBA images show not only a jet at different position angles, but in
addition the jet is sometimes double-sided \citep[and references therein,
but see \cite{2006smqw.confE..52D} for interpretations with pulsar wind model]{2012A+A...540A.142M}.  One-sided jet
structures are seen in blazars, because of their small angle with respect to
the line of sight, counter-jet emission is de-boosted below detection
sensitivity.  The switch in \target between a two-sided and a one-sided
structure indicates that precession bringing the jet to small angles with
respect to the line of sight.  The radio images confirm therefore a variation
of the angle between jet and line of sight and indicate that rather small jet
to line-of-sight angles are reached.  Information on periodicity came from 2006
astrometry \citep{2006smqw.confE..52D}.  \citet{2012A+A...540A.142M} showed as
the peak of the images, associated with the jet core, described an ellipse path
on the sky over 27-28 d, i.e., close to $P_2$.  In the following sections we present
new observations and we compare the astrometry data from 2006 and 2015 with the
predictions of a precessing jet model.

 \section{Observation and Data reduction}
We conducted 10-epochs of phase-referenced observations with the VLBA from 2015
July 24 to August 23, spanning one orbital cycle of the binary with roughly 3-d intervals. At each epoch observations spanned 4 hours. The angular
separation between \target and the calibrator, J0244+6228, is 1.3$^\circ$.  The
antenna switch time, i.e., one phase-referencing cycle of the calibrator
J0244+6228 and \target is 3 min.  The total on-source times for \target, J0244+6228 and
the astrometric check source, J0239+6005, were 2.0, 1.2 and 0.5 hours, respectively. 
The synthesized beams were 2.2$\times$1.0~mas at a position angle of
0$^\circ$ (East of North) for all epochs, except for epoch D for which the
MK and NL antennas were unavailable and the synthesized beam was
2.4$\times$1.0~mas at position angle of 51$^\circ$. We used identical
frequency setups as were used for the 2006 VLBA observations: continuum
emission at 8.4~GHz, was recorded with four adjacent dual circular
polarization intermediate freqency (IF) bands of 16~MHz and
correlated with 64 channels per IFs. 

The 2006 and 2015 datasets were calibrated using Astronomical Image Processing
System (AIPS)\footnote{AIPS is a NRAO software package to reduce radio
interferometric data that can be available from http://www.aips.nrao.edu}
together with scripts written in ParselTongue, a python interface
to AIPS and Obit \citep{2006ASPC..351..497K}.  We first corrected the data for
residual delays from Earth orientation parameters (EOP) and ionosphere (TECOR);
amplitudes were adjusted for digital sampling corrections (ACCOR), system
temperatures and antenna gains; manual phase-calibration was
applied to correct for delay offsets between sub-bands.  After these
procedures, the phase-reference source, J0244+6228, was imaged and
self-calibrated using the procedure Muppet
in DIFMAP \citep{1994AAS...185.0808P, 1997ASPC..125...77S}.  Then the 
self-calibrated J0244+6228 images of individual epochs were read back into
AIPS and served as a model to calibrate delay, rate and
phase with FRING. The structure of J0244+6228 appeared stable throughout
both the 2006 and 2015 sessions. The self-calibration solutions
were then interpolated to the \target and J0239+6005 scans; the
latter source was used as an astrometric check. Finally, phase-referenced
images were produced by IMAGR. The peak positions of \target, which is neither
Gaussian-like nor symmetric, were determined
from the brightest pixels (of size 0.05 mas$\times$0.05 mas) with IMSTAT
by following the approach of \citet{2012A+A...540A.142M}.

 \begin{figure}
 \includegraphics[width=8cm]{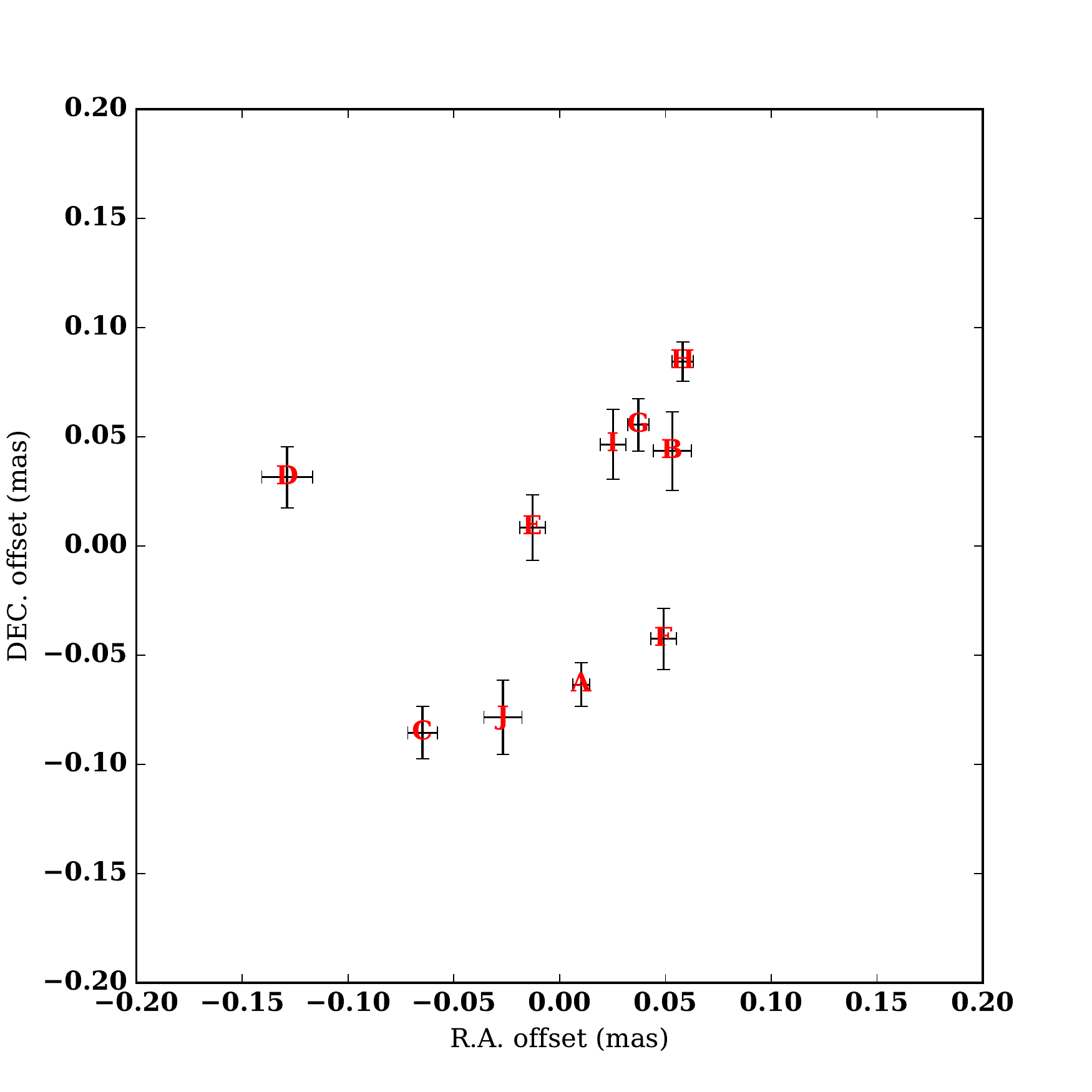}
 \caption{
 Astrometric accuracy of J0239+6005. Red characters denote epochs. The
 standard deviation is 0.05 mas. The error bars are formal position errors
 of J0239+6005 determined by 2-D Gaussian fitting of brightness distribution of
 the phase-reference images with AIPS task JMFIT. \label{fig-1}}
 \end{figure}

In Figure \ref{fig-1}, we present the astrometry obtained for the check source
J0239+6005 relative to the phase-reference source J0244+6228 from the 2015 VLBA
observations. We used the AIPS task JMFIT to fit positions and the formal
errors are $\sim$~0.008~mas and $\sim$~0.015~mas in R.A. and Dec. directions,
respectively.  The standard deviation of these positions are 0.05 mas in both
R.A. and Dec directions\footnote{The R.A. offsets in the Figure are
$\Delta$R.A.~$\times$~$\cos$(DEC).}. Table \ref{tab-1} lists the jet-core
positions of \target from the 2006 and 2015 sessions and the estimated orbital,
precessing and long beating phases. 

 \section{precession period}
 \begin{figure*}
 \includegraphics[width=18cm]{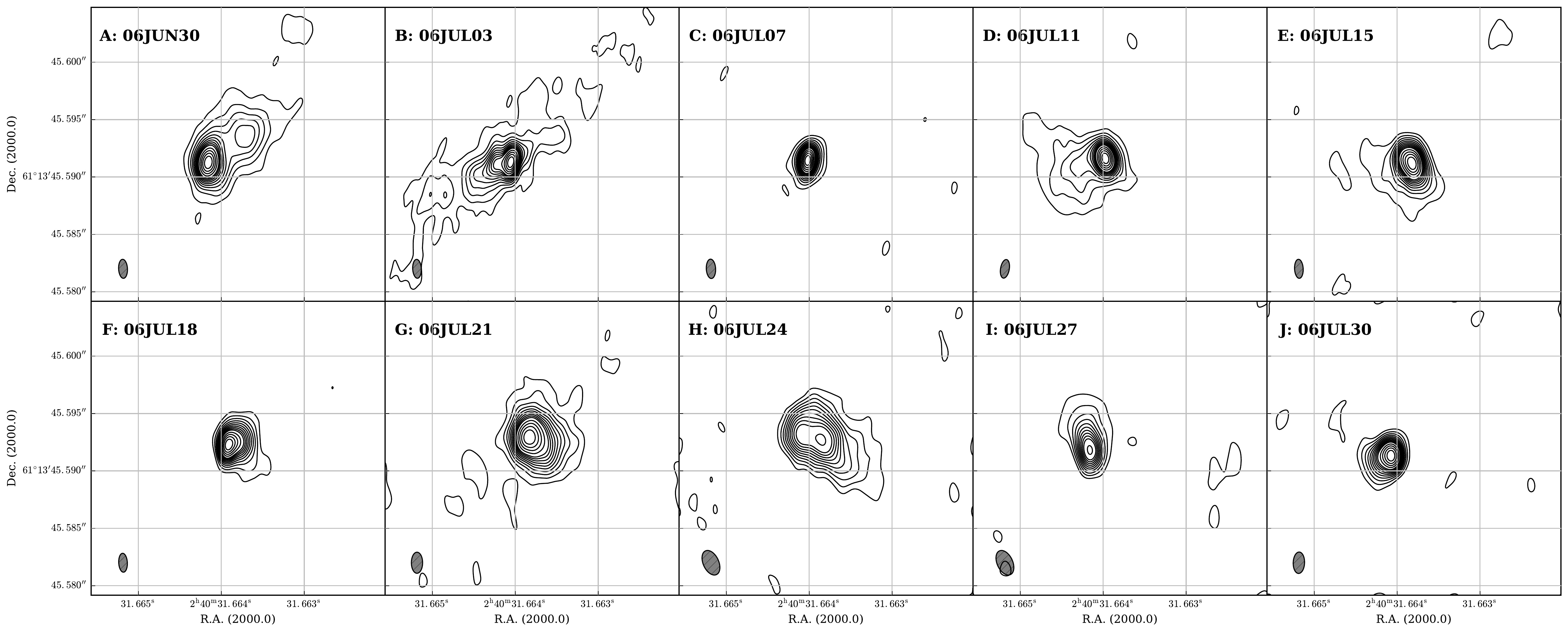}
 \includegraphics[width=18cm]{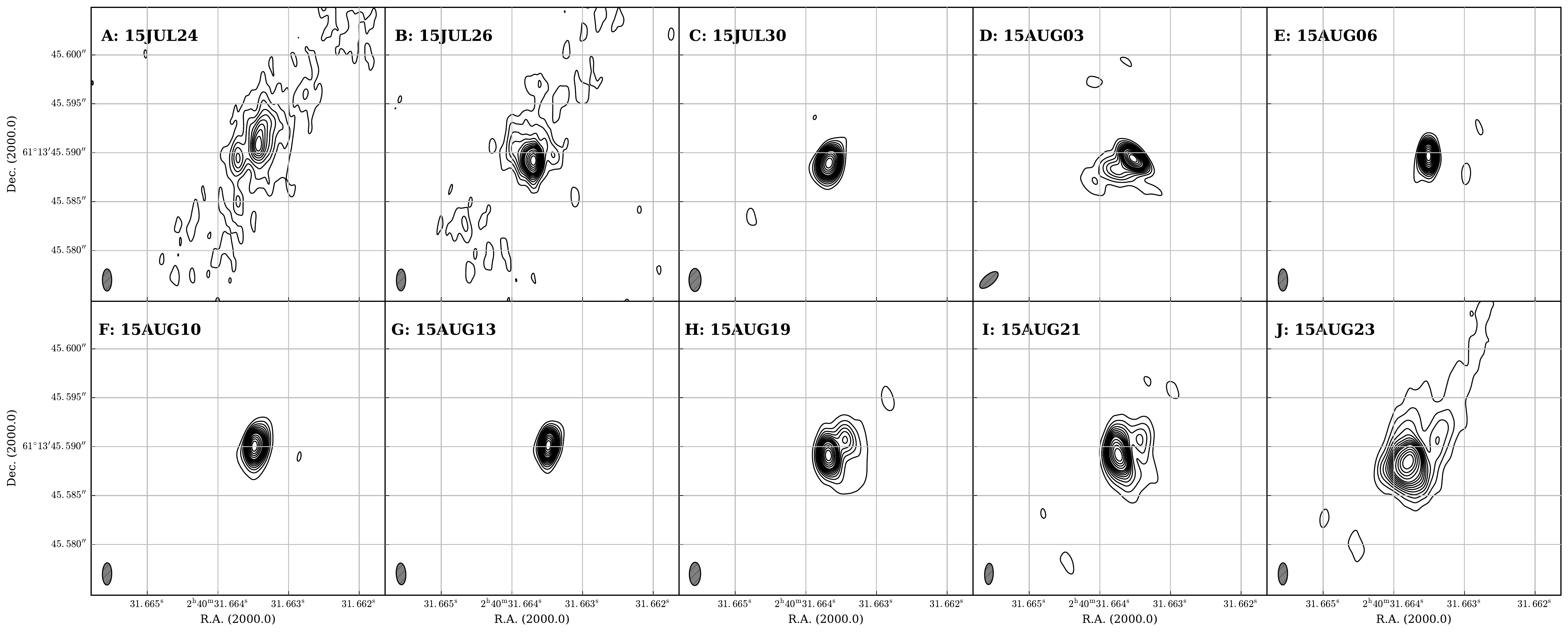}
 \caption{Phase reference images {of \target in} 2006 and 2015 observations. The
 observational dates are labeled in the top left corner. The hatched grey areas
 in the bottom left corner denote the synthesized beams. All contour levels start
 from 5$\sigma$, and increase with step of 5$\sigma$ (here $\sigma$ is the rms noise level of the
 image, which typically was $0.1$ mJy beam$^{-1}$).
 \label{fig-2}}
 \end{figure*}
 
In Figure \ref{fig-2}, we show the phase reference images of \target in 2006
and 2015, with observation dates labeled. Changes of jet position angles can be
clearly seen in both sessions. In most epochs, the jet appears one-sided;
however, in epochs of 06JUL03, 06JUL24 and 15JUL24, there are two-side
morphologies. This phenomenon can be explained by precession of the jet axis,
as illustrated in Figure 7 of \citet{2014A+A...564A..23M}, especially when the
angle between the line-of-sight and the jet axis is minimized and Doppler
boosting is maximized.

\begin{figure*}
 \includegraphics[width=18cm]{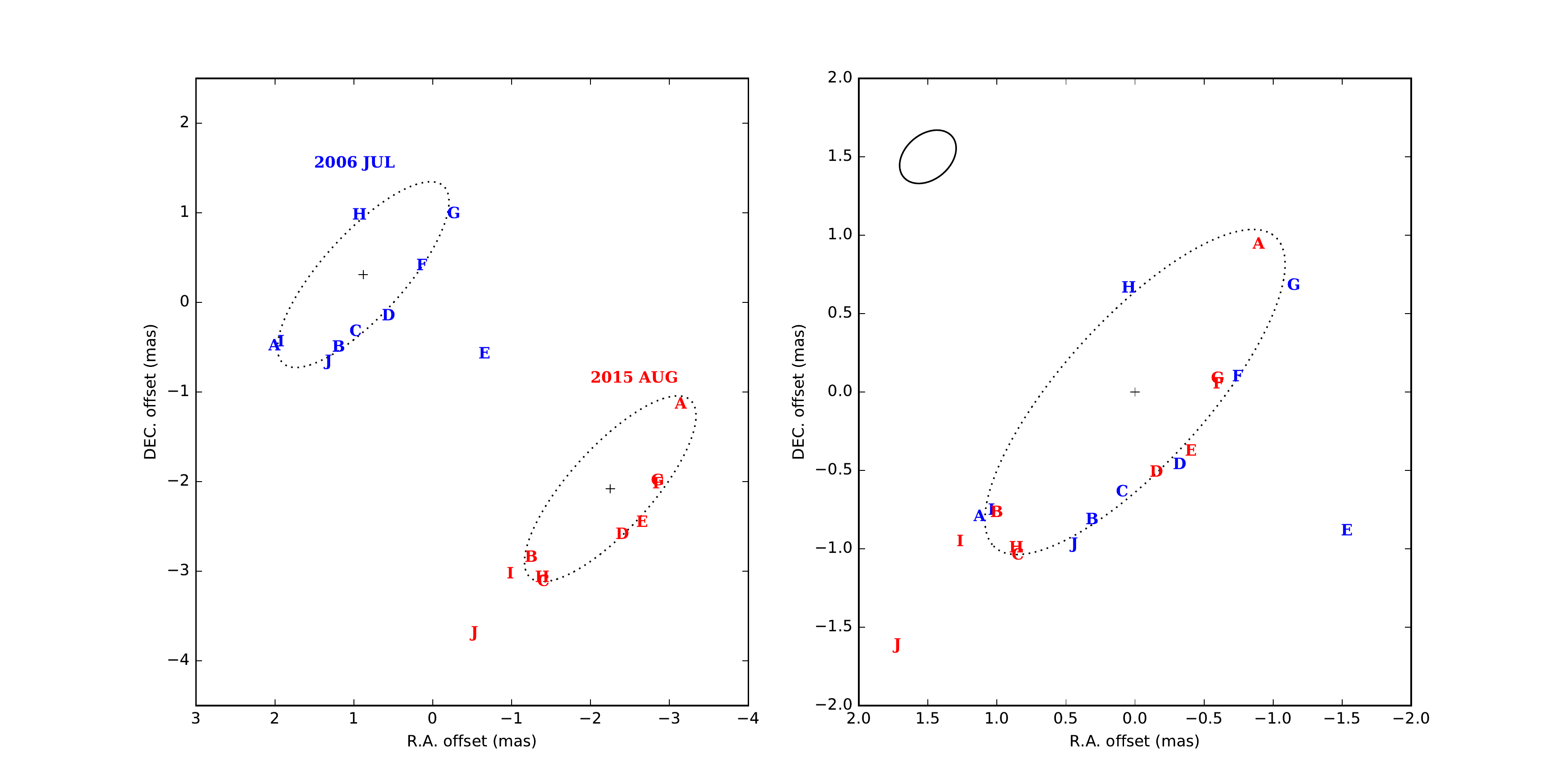}
 \caption{$Left~panel$: Astrometric results of 2006 and 2015 VLBA
observations, with parallax motions removed. Blue characters denote jet peaks
in 2006, and red characters denote jet peaks in 2015. The reference coordinate (zero point) is 02h40m31s.6645,
61d13m45s.594. 
 $Right~panel$: Same as left panel, but with centers of the two ellipses overlaid.
 The solid ellipse in the top left corner indicates the scale of the
 orbit, with a semimajor axis of 0.22 mas \citep{2012A+A...540A.142M}. 
 \label{fig-3}} 
 \end{figure*}

The apparent motions of the jet core can be attribute to four components: (1)
orbital motion of $\approx 0.2$ mas, (2) jet precession of $\approx 2.5$ mas,
(3) proper motion of $\approx 4$ mas in 9 years and (4) parallax changes of $\approx 0.2$
mas. Orbital motion and precession at the jet core's position will be discussed
in detail in Section 5. The apparent motion caused by the source's parallax can
be well modeled since its distance of 2~kpc is known. In the left panel of
Figure \ref{fig-3}, we show the astrometric results with the parallax component
removed, where the reference coordinates (ie, the zero point) are 02h40m31s.6645,
61d13m45s.594 (J2000). In both the 2006 and 2015 sessions, one can see the elliptical
trajectory. We align the two ellipses by fitting the jet-core positions shown in
the left panel of Figure \ref{fig-3} with seven parameters ($x_1$, $y_1$,
$x_2$, $y_2$, $a$, $b$, $\Theta$), where ($x_1$,$y_1$) and ($x_2$,
$y_2$) are centers of two ellipses, $a$, $b$ and $\Theta$ are the
long/short axes and the position angle of the long axis. 
(The best fitting values of these parameters, respectively, are 0.87$\pm$0.04 mas, 0.31$\pm$0.05 mas,
-2.23$\pm$0.07 mas, -2.08$\pm$0.03 mas, 1.39$\pm$0.05 mas, 0.48$\pm$0.11 mas, and -43.63$\pm$1.64 deg). 
Note, we have assumed an identical elliptical shape for the two sessions, 
with the only difference being the central positions. 
The right panel from Figure \ref{fig-3} shows the
relative positions of the jet cores referred to the elliptical centers in 2006 and
2015 sessions.  One can see that the 2006 and 2015 astrometric data can be
well modeled by the source proper motion (see Section 6) and two identical elliptical
trajectories caused by jet precession. 

We then determine the precession period by aligning the phase of these two
ellipses.  The ellipse in polar coordinates can be denoted as 

 \begin{equation}
 r = \frac{a \times b}{\sqrt{a^2 \sin^2 (\theta - \Theta)+b^2 \cos^2 (\theta - \Theta)}}
 \label{eq-a}
 \end{equation}
where the reference point is the elliptical center, the reference direction
is the east direction, $a$, $b$ and $\Theta$ are the long/short
axes and the position angle of the long axis. The polar coordinates $r$
and $\theta$ can be converted to the Cartesian coordinates $x$ and
$y$ using trigonometric functions

 \begin{figure*} 
 \includegraphics[width=18cm]{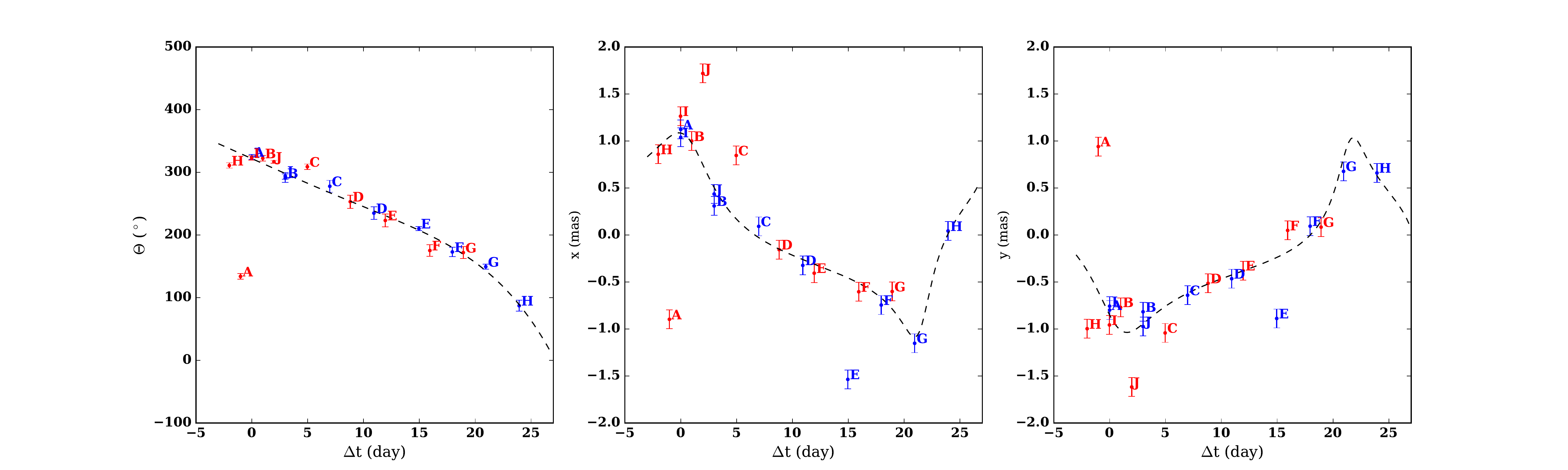}
 \caption{Left, middle and right panels are the position angle $\theta$, $x=\Delta$R.A. and $y=\Delta$Dec. 
  of jet core versus time, respectively. Blue/red characters and dots with error bars
denote 2006/2015 observations.
 \label{fig-4}} 
 \end{figure*}

 \begin{figure} 
 \includegraphics[width=9cm]{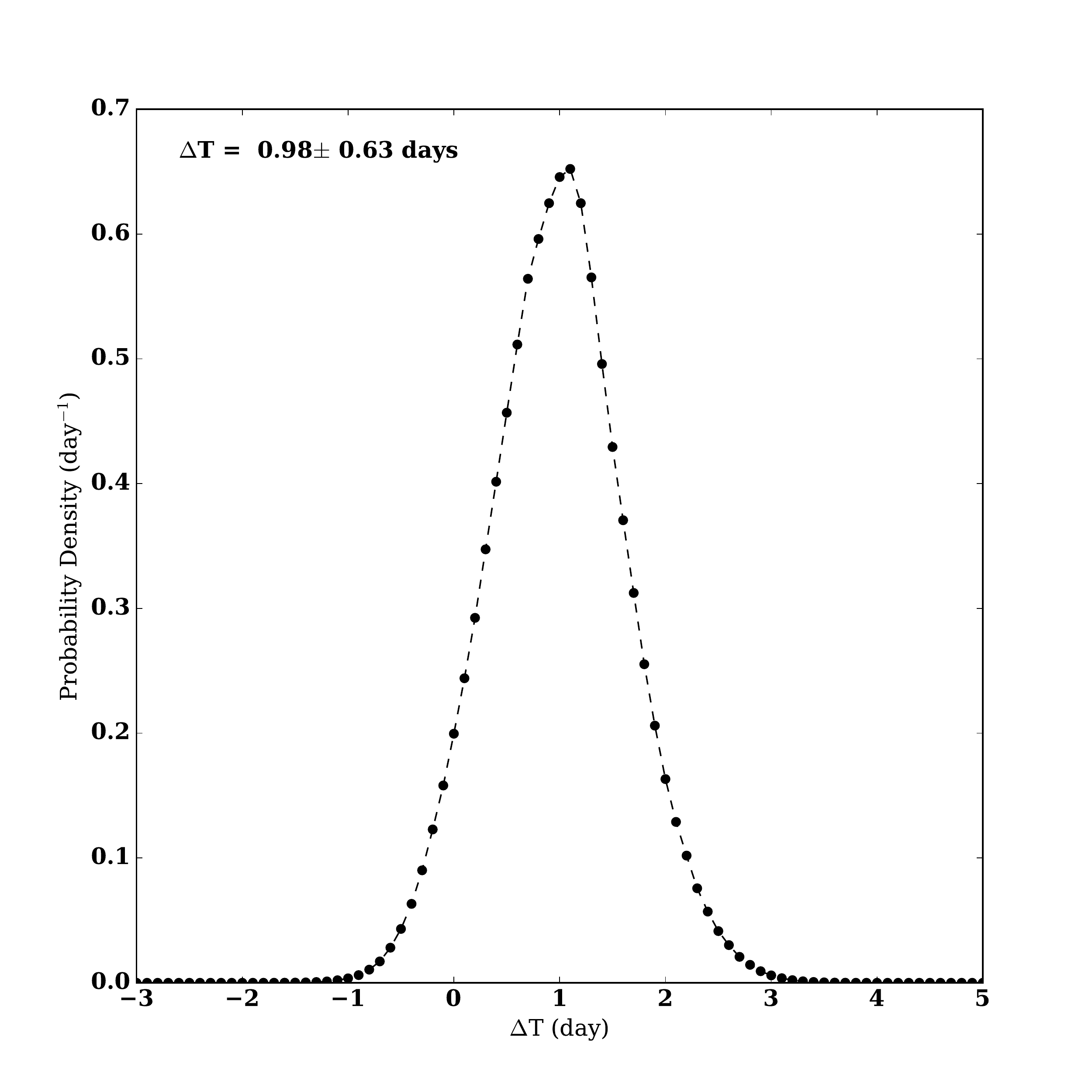}
 \caption{Probability density function for $\Delta$$T$. The PDF
  was calculated at steps of 0.1~d, shown with black dots connected by
  the dashed line.\label{fig-5}} 
 \end{figure}

 \begin{figure*} 
 \includegraphics[width=5cm]{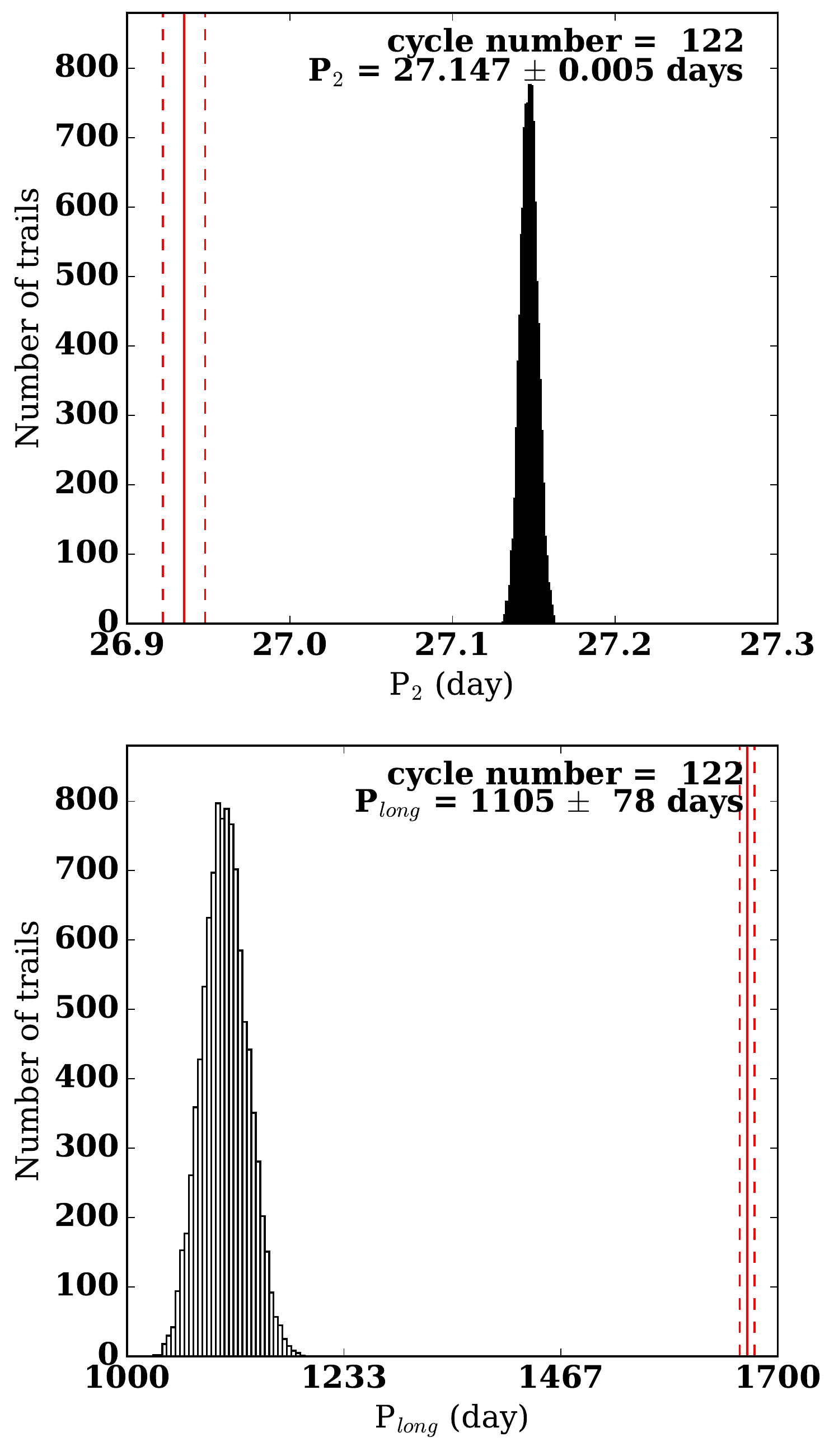}
 \includegraphics[width=5cm]{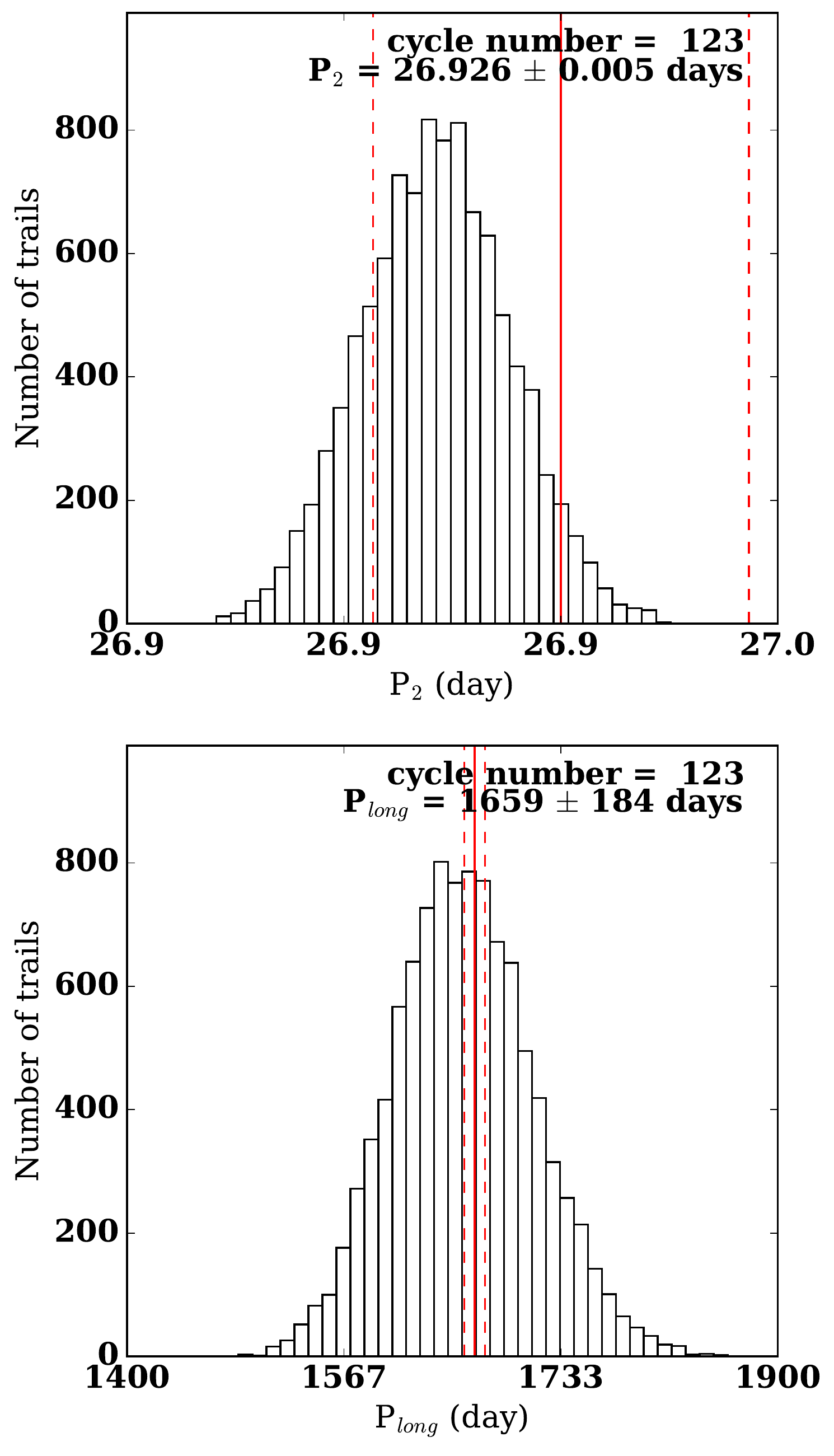}
 \includegraphics[width=5cm]{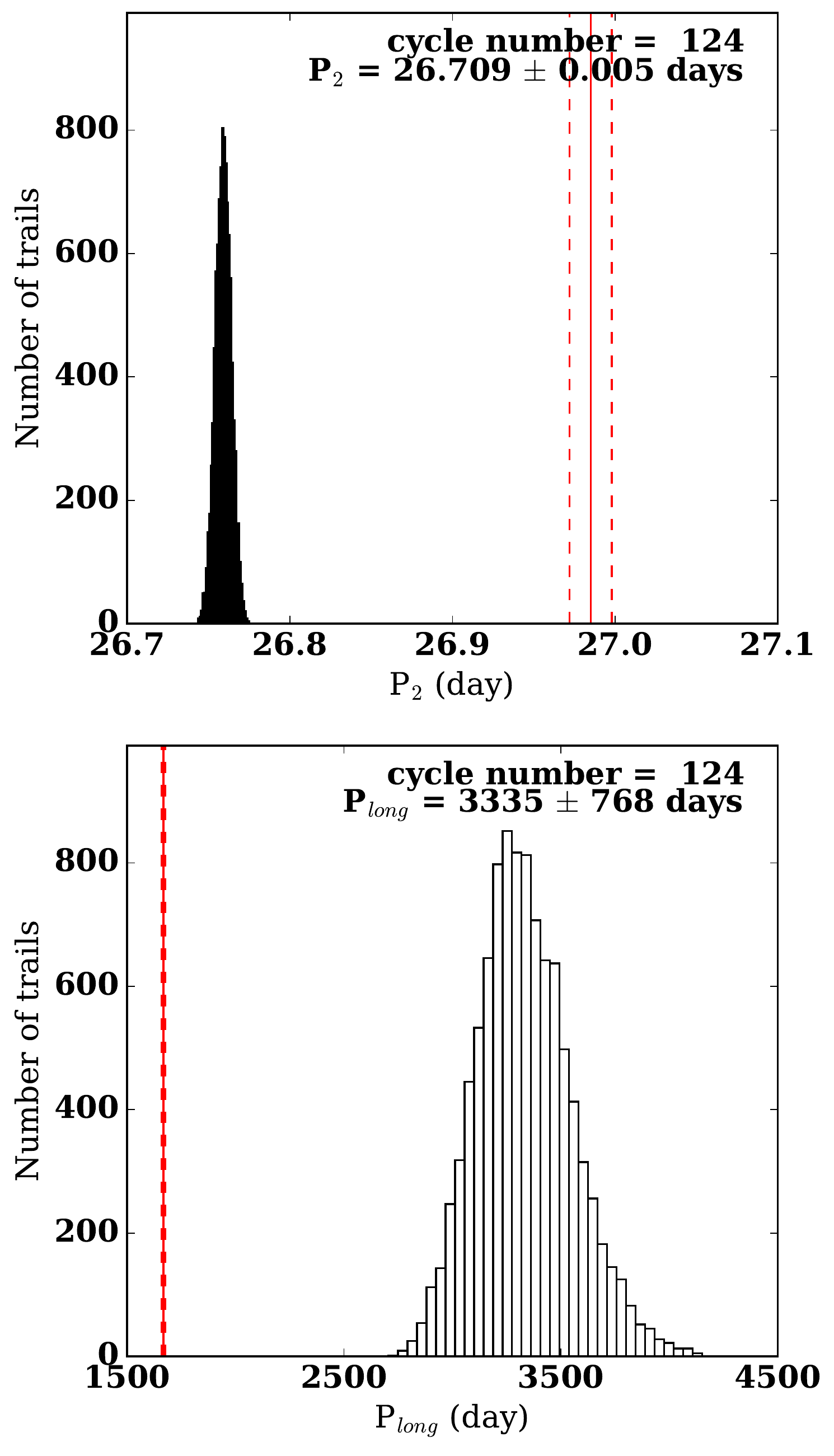}
\caption{\label{fig-6}$Top~panels:$ Histograms of 
$P_2$ estimated via Monte Carlo simulation, by adopting
$\Delta$$T$~=~0.98$\pm$0.63~d, with cycle number~=~122, 123 and 124 (top left,
top middle and top right panels). We ran 10000 trials to estimate the value and
1$\sigma$ error of  $P_2$.  $Bottom~panels:$
Histograms of $P_{\rm long}$ estimated with relationship of
$P_{\rm long}$~=~$P_{1}$$\times$$P_{2}$/($P_{2}$-$P_{1}$), by adopting
$P_1$~=~26.496~$\pm$~0.013~d \citep{2016A+A...585A.123M},
and $P_2$ values from top panel estimations. Red solid
and dashed vertical lines in top and bottom panels denote values and 1$\sigma$
errors of $P_2$ and $P_{\rm long}$ estimated by
\citet{2016A+A...585A.123M} and \citet{2002ApJ...575..427G} with Lomb-Sargle
method, respectively.
\label{fig-6}} 
\end{figure*}

In Figure \ref{fig-4}, we present $\theta$, $x$ and $y$ vary with time, where
$\Delta$$t$~=~$t$ - $t_{\rm ref}$. We assume an identical precessing phase at the
reference time, $t_{\rm ref}$. For 2006 and 2015 sessions, we set
$t_{\rm ref}$~=~$t$[2006-A] and $t_{\rm ref}$~=~$t$[2015-B]-$\Delta$$T$,
respectively. Here the $\Delta$$T$ is the parameter that we need to estimate. 

In the left panel of Figure \ref{fig-4}, the dash line is an empirical
relationship between $\Delta$$t$ and $\theta$, that is a monotonic polynomial
fitted using data of 2006 sessions.
 \begin{equation}
 \theta{({\Delta}t)} = a + b\times{\Delta}t + c\times{\Delta}t^3 + d\times{\Delta}t^5
 \label{eq-c}
 \end{equation}
Here we used an odd order polynomial (i.e., with diagonal symmetry rather 
than mirror symmetry for an even order).  The 5th order polynomial was adopted, 
as it produced much smaller scatter than a 3rd order polymomial and 
a 7th order polynomial did not yield significant improvement.
The dash lines in the middle and right panels then can be determined with
equation (\ref{eq-a})-(\ref{eq-c}).

The best $\Delta$$T$, can be estimated as the maximum of the probability
density function (PDF), which is defined as 

\begin{equation}
Prob \propto \prod_{i=1}^{N}\frac{1}{\sigma\sqrt{2\pi}}e^{-\Delta_i^2/2\sigma^2}
\label{eq-d}
\end{equation}
where, 
\begin{equation}
\Delta{_i^2} = (x_{i}[obs]-x_{i}[model])^2+(y_i[obs]-y_i[model])^2
\label{eq-e}
\end{equation}
are residuals (data minus model); $\sigma$ was estimated from the post-fit residuals from

\begin{equation}
\sigma^2 = \sum_{i=1}^{N}\frac{\Delta_i^2}{N}.
\label{eq-f}
\end{equation}
In Figure \ref{fig-5}, we show the PDF for $\Delta$$T$, from which we
estimate $\Delta$$T$~=~0.98$\pm$0.63~d. The PDF is calculated with a step of 0.1~d.

 The time interval between 2006-A and 2015-B is 3312.9174 d, take into
 account $\Delta$$T$~=~0.98~$\pm$~0.63 d, the time interval between
 these two precessing cycles are 3311.9 $\pm$ 0.6 d. We can determine the
 precessing period, once the cycle number between these two periods is known.
 Meanwhile, if we assume the relation  
 $P_{\rm long}$ = $P_1$$\times$$P_2$/($P_2$-$P_1$)
 holds \citep[e.g.,][]{2016A+A...585A.123M},  then we can calculate $P_{\rm long}$,
 once $P_1$ and $P_2$ are given. 
 In upper and lower panels of Figure \ref{fig-6}, we
 present $P_{2}$ and $P_{\rm long}$ estimated via Monte Carlo simulations, assuming there
are 122, 123 and 124 cycles over the time interval of 3311.9 $\pm$ 0.6 d.
In Figure \ref{fig-6}, the red solid/dashed lines
 indicate $P_2$~=~26.935~$\pm$~0.013~d \citep{2016A+A...585A.123M}, and
 $P_{\rm long}$~=~1667~$\pm$~8~d \citep{2002ApJ...575..427G}. It can be seen 
 that only 123 cycles can produce consistent  $P_2$ and
 $P_{\rm long}$ values.  With
 the value of 123 cycles, we determined an accurate $P_2$,
 \begin{equation} 
 P_2 = (3311.9~\pm~0.6)/123 = 26.926~\pm~0.005~d~~. 
 \label{eq-g}
 \end{equation}

 \section{Modeling the astrometry of the jet core} 
The emissions of \target from $\gamma$-rays, X-rays, optical/infrared, and radio wavelengths have
 been modeled by several authors \citep{1992ApJ...395..268T,
 1995A+A...298..151M, 2006A+A...459L..25B, 2007A+A...474...15R,
 2014A+A...564A..23M} in the context of accretion onto a compact object along an eccentric orbit.  
 Observational evidence, especially from measurements of the radio spectral index
 \citep{2009ApJ...702.1179M} and a high energy double-peak light curve
 \citep{2016A+A...595A..92J} favors a \mqso\ rather than a pulsar wind
 origin \citep{2006smqw.confE..52D}. \citet{2014A+A...564A..23M} developed a model of a precessing conical
 jet which emits synchrotron radiation to explain the radio light curve.  
 In this section, we integrate their radiation transfer model in order to simulate observations 
 on the sky plane.   For an optically thin
 jet, the maximum of the emission is at the jet base,  while for an optically
 thick jet the maximum will be displaced down the jet where
 optical depth unity is achieved.  The
 displacement of the observed radio peaks on the sky plane is due to 
 the emitting plasma changing position owing to the
 orbital motion of the compact object around the primary star and to the jet
 precession.  In this section, we first examine the two effects separately and
 then we derive the full jet motion as it appears on the sky.
 
 \subsection{The orbital motion}
\begin{figure}
 \centering
 \includegraphics[width=8cm]{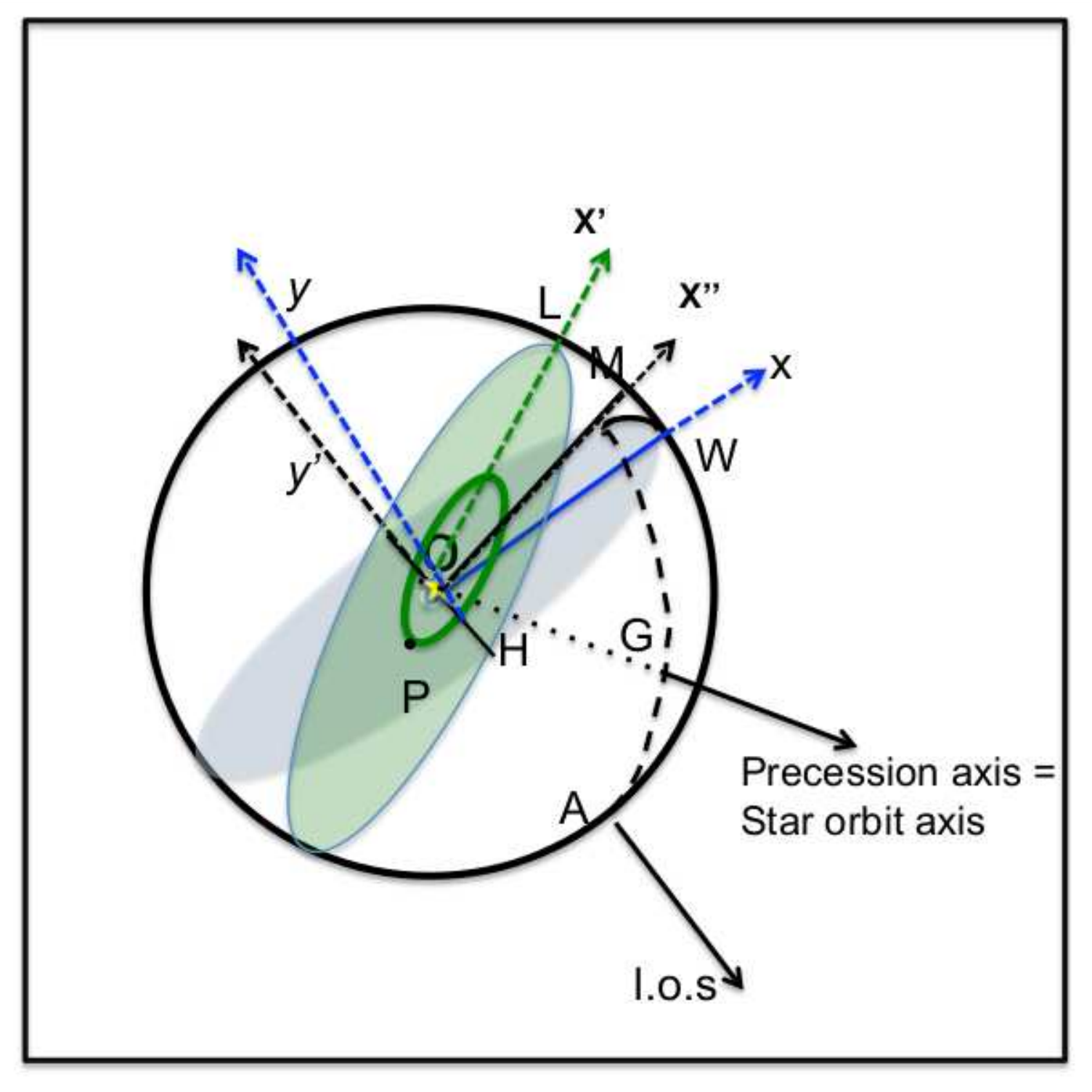}
 \caption{Geometry of the orbit. The real  orbital plane is in green; the
 grey plane is perpendicular to the observer.  The center of mass of the
 system is located at $O$; $\hat{WOL }~=~\zeta$ is the angle between the orbital 
 plane and the plane perpendicular to the line of sight; 
 $\omega~=~90^o-\hat{POH}$ and $\hat{MOW}$~=~$w$.  } 
 \label{fig-7}
 \end{figure}

 In \cite{2014A+A...564A..23M} model,  the base of the jet is anchored to the
 compact object and, hence, follows the compact object along its orbit;
 this path is drawn as a green ellipse in Fig.~\ref{fig-7}.  The orbit plane
 forms an angle $\zeta$ with respect to the plane perpendicular to the 
 line-of-sight.  
 
The system of reference in the orbital plane is at the center of mass of the
system i.e. the point $O$.   The $y'$ axis  is defined by the intersection of
the orbital plane and the plane perpendicular to the line-of-sight, with the
$x'$ axis perpendicular to the $y'$ axis at $O$.  In this way a rotation of the
system [$x'$,$y'$] around [$y'$] by an angle $\zeta$ defines the system of
reference [$x''$,$y'$] in the plane perpendicular to the line-of-sight.  In
order to parameterize the orientation of the ellipse in the orbital plane we
introduce the angle $\omega$, which defines the ellipse rotation with respect
to the Cartesian system [$x'$,$y'$] previously defined. See Fig.~\ref{fig-7}
for angle definitions. 

 The compact object moves at a distance $\rho$ from the origin of coordinates.
 In the plane of motion, the vector radius can be expressed in terms of the
 ellipse semi-major axis, $a$, ellipse eccentricity, $e$, and the angle
 $\theta$ (which is zero when the vector radius points toward apastron) as

 \begin{equation}
 \rho(\theta) = {a~(1-e^2)\over 1-e ~\cos\theta}
 \label{eq-2}
 \end{equation}
 The components of the vector radius in the
 [$x'$,~$y'$] orbital plane, for an 
 ellipse rotation of an angle $\omega$ , are
 \begin{equation}
 \rho_{x'} = {a~(1-e^2) \cos(\theta+\omega)\over 1-e~ \cos(\theta)}
 \label{eq-3}
 \end{equation}
 \begin{equation}
  \rho_{y'} = {a~(1-e^2) \sin(\theta+\omega)\over 1-e ~\cos(\theta)}
 \label{eq-4}
  \end{equation}

In the plane perpendicular to the line-of-sight, owing to a rotation of an angle
 $\zeta$ around [$y'$], these components become
 \begin{equation}
 x''= \rho_{x'} ~\cos \zeta  \\\    y' = \rho_{y'}.
 \label{eq-5}
 \end{equation}
and, in the system of reference [$x$, $y$],
 \begin{equation}
 x_{orbit} = x''\cos w - y' \sin w
 \label{eq-6}
 \end{equation}
 \begin{equation}
 y_{orbit} = y' \cos w +x''  \sin w
 \label{eq-7}
 \end{equation}
where  $w~=~\hat{MAW}~=~\hat{MOW}$ in Fig.~\ref{fig-7} is a free angle, 
to be determined by the fit (see Sec. 5.3).

 \subsection{The jet precession motion}

\begin{figure}
 \centering
    \includegraphics[height=7.8cm]{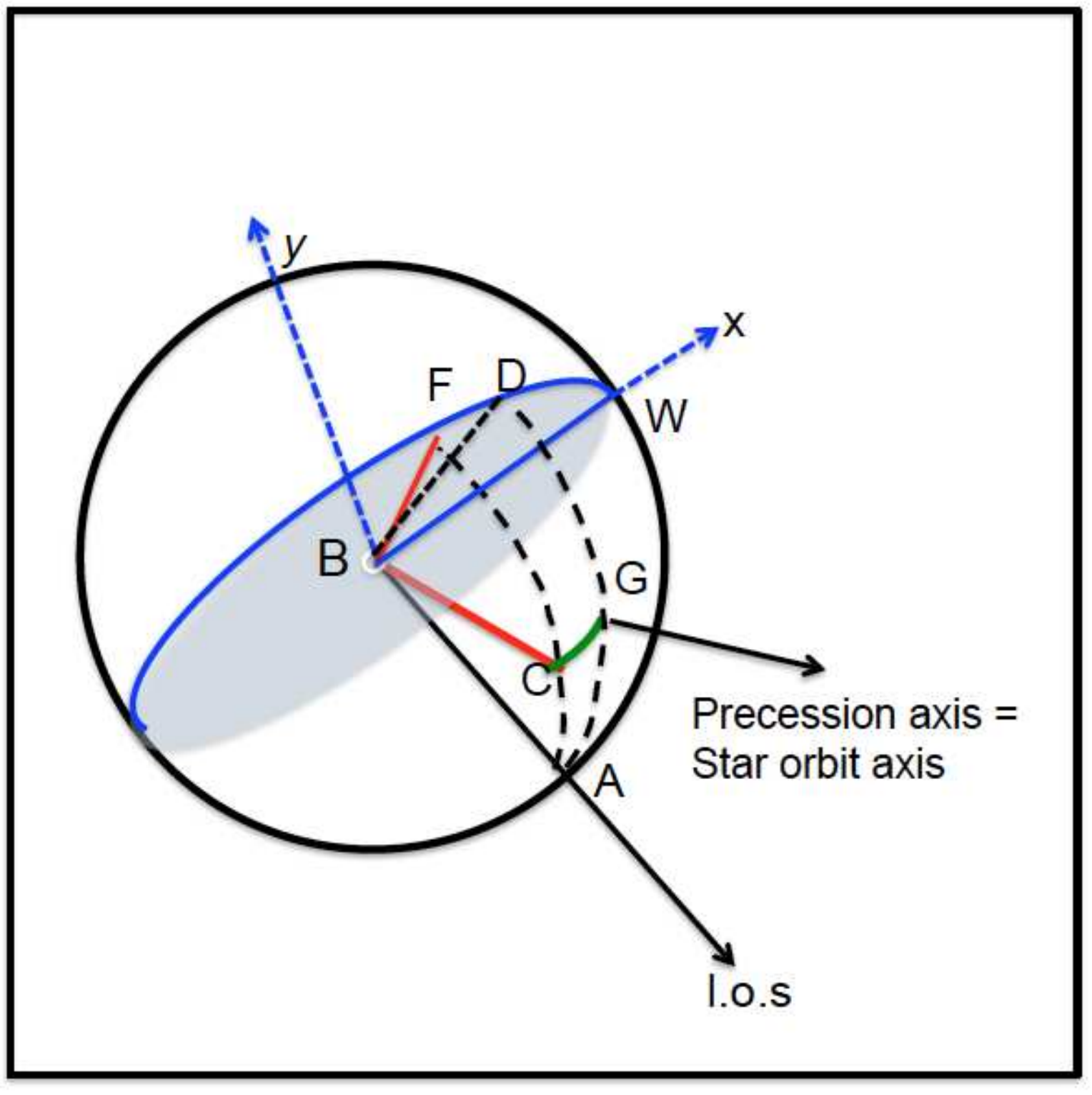}\\
 \caption{
 Geometry of the precessing radio jet. $BC$ is the distance of the jet base
($x_0$) from the compact object (located in $B$);  $BF$
is the  projection of $BC$  on the plane perpendicular to the line of sight; $CA$~=~$\eta$;
$CG$~=~$\psi$; $AG$~=~$\zeta$;  $\hat{CGA}~=~\Omega$; 
 $\hat{ FBD}$~=~$\hat{CAG}$~=~$\delta$; 
 $\hat{DAW}$~=~$\hat{DBW}$~=~$w$ free angle to be determined.} 
  \label{fig-8}
 \end{figure}

We assume the jet geometry and properties as described in \cite{2014A+A...564A..23M}.
In Fig.~\ref{fig-8} the base of the jet is at point $C$.
The jet base will be projected at a distance $BF~=~BC~sin \eta~=~x_0~sin \eta$,
where $\eta$ is the angle the jet makes with the line-of-sight
and $x_0$ (i.e., $BC$ in Fig.\ref{fig-8}) is the position of the jet base.
Hence, for a thin jet,the components of the emitting region
with respect to the plane perpendicular to the line of sight are 
 \begin{equation} x_j~=~x_0 \sin \eta~\cos(\delta+w) \\ y_j~=~x_0 \sin \eta~ \sin(\delta+w)~~, 
 \label{eq-8}
 \end{equation} 
where the angle $\delta$ can be derived from the spherical triangle $ACG$ (see Fig.\ref{fig-8}) as
 \begin{equation}
 \sin \delta~ \sin \eta~=~\sin \Omega ~\sin \psi
 \label{eq-9}
 \end{equation}
 \begin{equation}
 \cos \delta~ \sin \eta ={ \cos \psi -\cos \eta~ \cos \zeta \over \sin \zeta}~,
 \label{eq-10}
 \end{equation}
and the angle $\Omega$, which changes because of precession, is defined as  
 \begin{equation}
 \Omega = 2 \pi t/P_{2}~.
 \label{eq-11}
 \end{equation}

 \subsection{The total motion: results}
 
 Since the jet is anchored to the accretion disk of the compact object,  the generic coordinate of the
 jet emitting plasma with respect to the assumed center of coordinates of
 Fig.\ref{fig-7} will be

 \begin{equation}
 X = l~x_j+x_{orbit}  \\\     Y = l~y_j+y_{orbit}
 \label{eq-12}
 \end{equation}
 where $l$  specifies the position along the jet of the peak radio flux;
 For an optically thin jet $l~=~1$, while for an optically thick jet 
 $l$ is the position of optical depth unity.
 The value of $l$ needs to be computed at each orbital phase in order  
 to derive the correct coordinates ($x$,$y$) of the radio emission.

\begin{figure*}
\includegraphics[width=7cm]{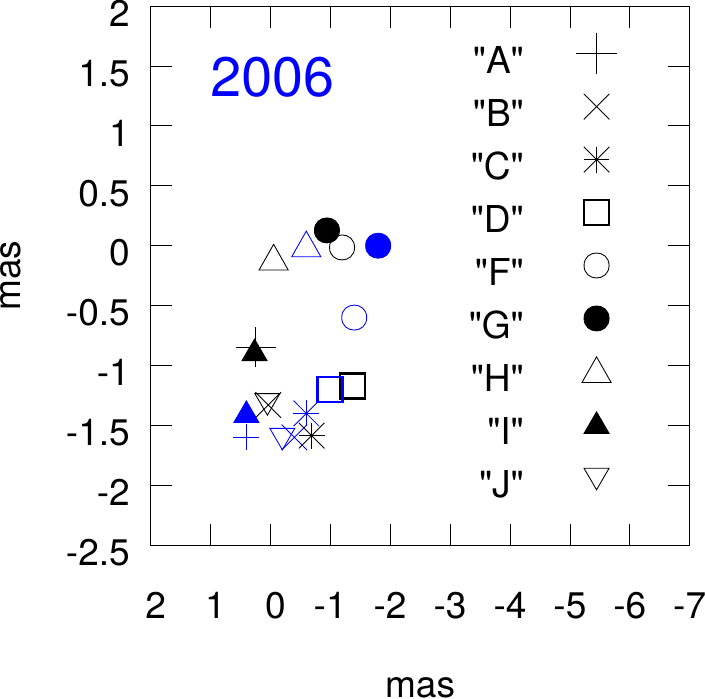}
\includegraphics[width=7cm]{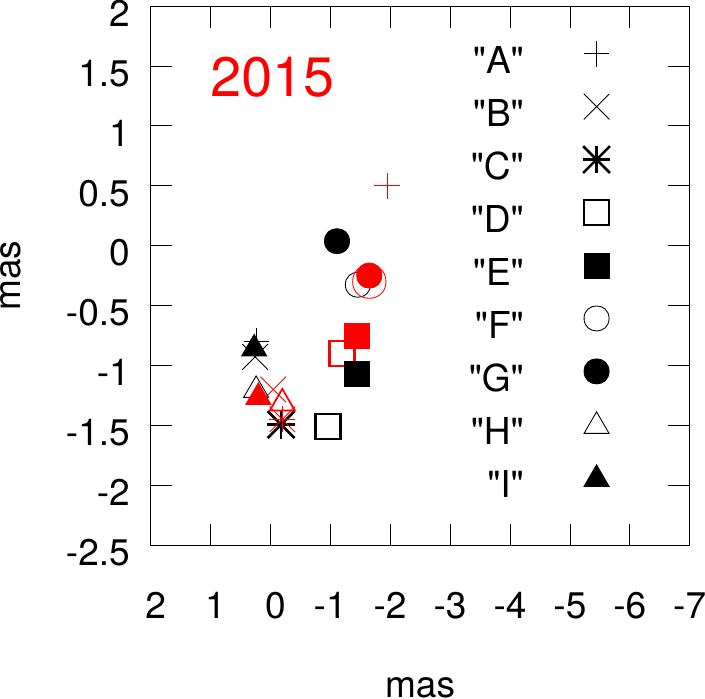}\\
\includegraphics[width=7cm]{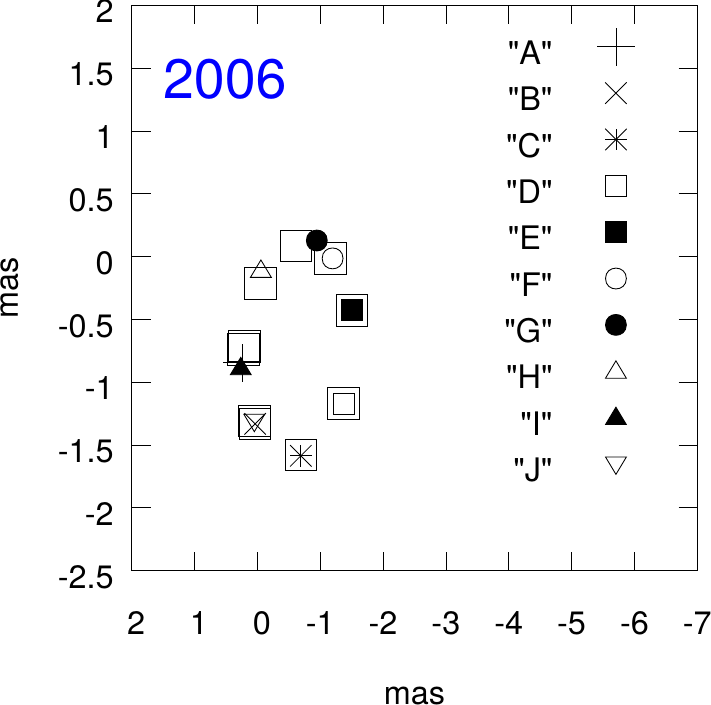}
\includegraphics[width=7cm]{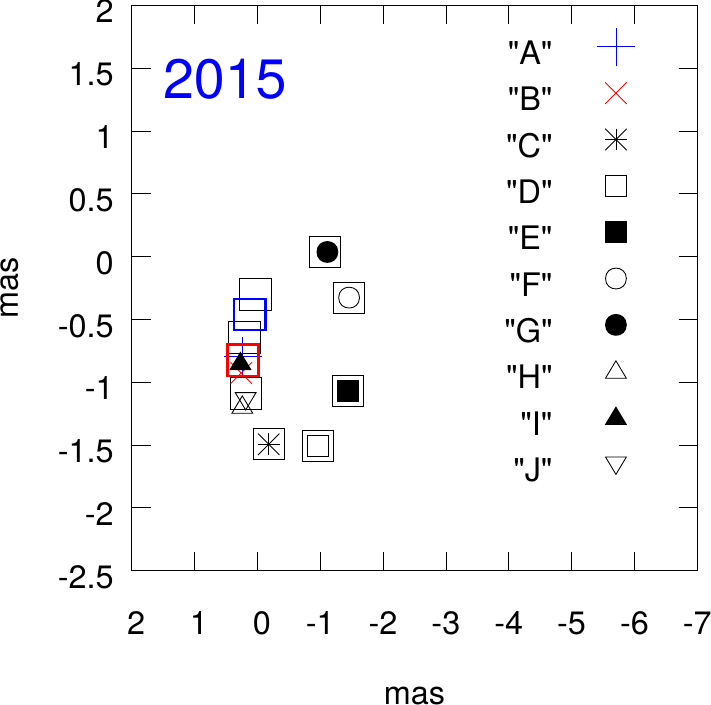}\\
\includegraphics[width=7cm]{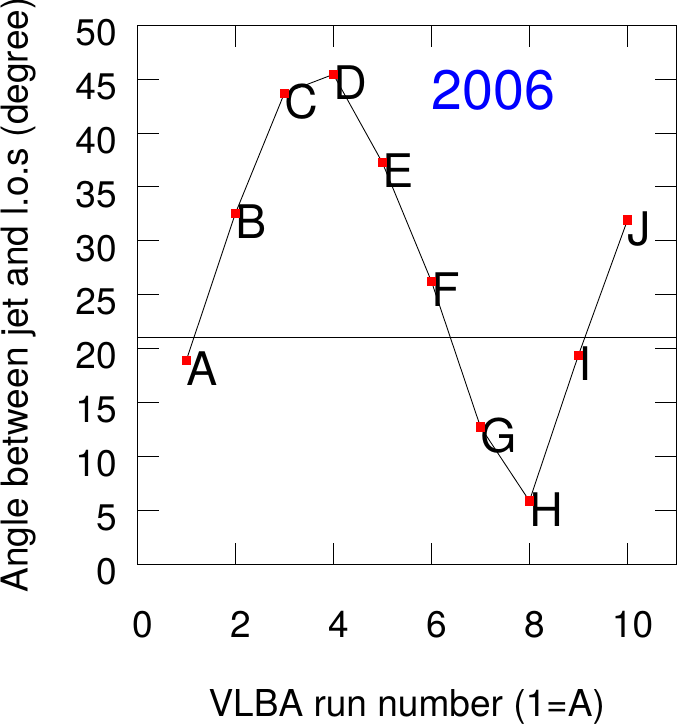}
\includegraphics[width=7cm]{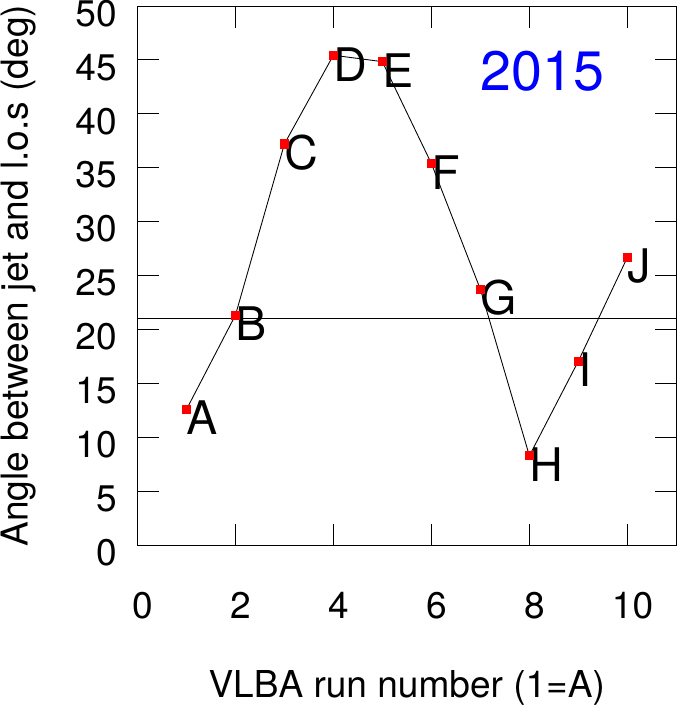}
\caption{
Top-panels: Model-data comparisons.
 The blue and red points are from VLBA astrometry in 2006 and 2015, respectively.
The same symbols in black are used for the model data.
 Middle  panels: 
 Comparison  of positions of the jet base (squares) and jet peak.
Bottom panel: angle between the jet and the line of sight (see Section 5.3).
\label{fig-9}
}
\end{figure*}

 The resulting positions for the model astrometry are shown in Fig.
~\ref{fig-9}.  We use a physical model, i.e., the number of relativistic
electrons injected every period $P_1$ in a conical jet at a particular orbital
phase, as found by \citet{2014A+A...564A..23M}, who fitted 6.7 yr of Green Bank 
Interferometer radio data. With astrometric position, we can now better constrain  
the geometrical parameters of the model;
i.e.,  $\psi$, the jet angle to its precession axis and $\xi$, the jet opening
angle.  The angle $\zeta$ is set to 25 degrees, which is the measured angle  
of the rotation axis of the Be star
\citep{2006PASJ...58.1015N}, assuming the star spin axis is parallel to the orbital axis.
The parameter $w$, that translates and rotates the trajectory in space without 
changing its shape is $w$~=~230 degrees.
The model for the astrometric data shown in Fig. ~\ref{fig-9} result in a jet angle 
to its precession axis of $\psi~=~21$ degrees and a jet opening angle $\xi~=~6$ degrees.  
We note that these two angles for \target are similar to those of 
the well-studied precessing jet system SS~433; SS~433 has a half
precessing cone angle of 20 deg and a jet opening angle of 5 deg 
\citep{1979ApJ...233L..63M, 1999A+A...348..910P}.

Since our physical model assumes a steady jet, any transient emission, e.g.,
observation E in 2006 \citep{2012A+A...540A.142M}, cannot be used in the fits.
For all other observations in 2006 the data and model overlap to within
$\pm3\sigma$.  The same occurs for 2015, except for observation A.  The model
is able to reproduce the main feature of the observed ellipse, i.e. the
non-uniform motion of the core with time, as seen in the middle panels of Fig.
~\ref{fig-9}.  For the 2006 observations, which were taken regularly every 3--4
d,  the jet base (large squares) follows a regular path.  The positions of
the core (at $\tau~=~1$) for A, H, I and G are displaced by the jet base and just
overlap with the edge of their related squares.  The same occurs in the 2016
session, where the predicted core positions for A, B, H, and I are offset from
the observed peaks.  As shown in the bottom panels of Fig. ~\ref{fig-9}, for
these observations the angle between jet and the line of sight is small, below
21 degrees.  Possibly the longer path of the radiation within the jet causes
the higher optical depth  \citep[see Fig. A1 in ][]{2014A+A...564A..23M}

 \section{Proper motion and 3D motion}

 Our observations also provide an accurate proper motion for \target as is
 evident in Figure \ref{fig-3}.  We find eastward and northward motions of
 $\mu_\alpha$~=~$-$0.150~$\pm$~0.006 mas~yr$^{-1}$,
 $\mu_\delta$~=~$-$0.264~$\pm$~0.006 mas~yr$^{-1}$, base
 on a time span of 3311 d between the 2006 and 2015 observations. 
 For comparison, the proper motions of \target measured by $Hipparcos$
 \citep{2000A+A...360..391H} and $GAIA$ \citep[Data Release
 1,][]{2017AA...599A..50A} are [0.62~$\pm$~1.95, 1.63~$\pm$~1.75]~mas~yr$^{-1}$
 and [-0.354~$\pm$~0.267, -0.077~$\pm$~0.211]~mas~yr$^{-1}$, respectively.
 \citet{2003AJ....126..484B} measured the proper motions of [0.97~$\pm$~0.26
 -1.21~$\pm$~0.32]~mas~yr$^{-1}$ with Very Large Array (VLA) data.  
 Except for the Boboltz et al values, previous estimates are consistent with 
 our results.  But, of course, our measurements have orders of magntitude better
 accuracy.
 
 Assuming a distance of 2.0~$\pm$~0.3~kpc and $V_{LSR}$~=~41.4~$\pm$~10 km~s$^{-1}$
 \citep{2009ApJ...698..514A}, we can estimate the full space velocity of the system.
 Adopting the distance to the Galactic center of 8.34 $\pm$ 0.16 kpc, 
 a \citet{1996MNRAS.281...27P} universal rotation curve with a
 circular rotation speed at the Sun of 240~$\pm$~8~km~s$^{-1}$, and solar motion 
 components of $U_\odot$~=~10.7~$\pm$~1.8, $V_\odot$~=~15.6~$\pm$~6.8~km~s$^{-1}$ 
 and $W_\odot$~=~8.9~$\pm$~0.9~km~s$^{-1}$ from \citet{2014ApJ...783..130R},
 we find peculiar (non-circular) of motion components for \target  of 
 $U$~=~10.8~$\pm$~8.8~km~s$^{-1}$, 
 $V$~=~$-$10.3~$\pm$~5.9~km~s$^{-1}$, 
 $W$~=~5.2~$\pm$~0.4~km~s$^{-1}$.  
 This translates to a speed of only 16~km~s$^{-1}$.
 
Recent analysis of $swift$ X-ray data suggest that \target is
a black hole \citep{2017arXiv170401335M}. Theoretically, stellar mass black
holes might be formed through different channels, i.e, with or without natal
kicks and with or without supernovae (SN) explosion. For example, GRO~J1655-40,
is a runaway black hole system with a peculiar speed of 112~$\pm$~18~km~s$^{-1}$
\citep{2002A+A...395..595M}.  It may be formed after an SN explosion created a
neutron star, followed by fall-back of ejected envelope and a secondary
collapse \citep{1999Natur.401..142I, 2002A+A...395..595M, 2003Sci...300.1119M}.
Alternatively, the black hole may be formed directly with a large natal kick
\citep{2012MNRAS.425.2799R}. In contrast, the peculiar motions of the well
studied black hole X-ray binaries, Cygnus X-1 and GRS~1915+105, have been
measured to be only $\approx20$ \kms\
\citep{2011ApJ...742...83R,2014ApJ...796....2R}.  Cygnus X-1 is believed to
have been formed through direct collapse of a massive star
\citep{2001ApJ...554..548F, 2003Sci...300.1119M}.  The similarly low peculiar
motion (16~\kms) of \target measured here suggest that it may also contain a
black hole formed through direct collapse of a massive star.
 
 \section{Conclusions}
 
 Using multi-epoch VLBA observations of \target in 2015, we were able to
 map the elliptical trajectory its radio emission.  The
 agreement between these maps and those from previous observations in 2006 
 suggests that the radio jet is stable over the nine year interval. 
 We then aligned the precessing phase and estimated the
 precession period to $P_2~=~26.926~\pm~0.005~d$, with a physical model
 that takes into account orbital motion, jet precessing and radiative transfer.
 In addition, the long time span between observations allow
 us to determine an accurate proper motion and, then, the full space motion of \target. 
 We find a small peculiar motion of 16~km~s$^{-1}$ for the system, which
 favors a black hole formation by direct collapse instead of a supernova explosion.
 
\section*{Acknowledgements}
\addcontentsline{toc}{section}{Acknowledgements}
This work has made use of data from the European Space Agency (ESA)
mission {\it Gaia} (\url{https://www.cosmos.esa.int/gaia}), processed by
the {\it Gaia} Data Processing and Analysis Consortium (DPAC,
\url{https://www.cosmos.esa.int/web/gaia/dpac/consortium}). Funding
for the DPAC has been provided by national institutions, in particular
the institutions participating in the {\it Gaia} Multilateral Agreement.
We would like to thank Prof. X.D. Li from the Nanjing University for useful
discussions.  We would like to thank the referee, Dr Benito Marcote, who
carfully read the original draft and share valuable comments and suggestions
that greatly imporve the quality of this paper.

{\it Facilities:} VLBA

\newpage

\begin{table*}
\caption{Jet core positions and estimated orbital, precessing and beating phases for 2006 and 2015 sessions
\label{tab-1}}

\begin{tabular}{cccccccc}
\hline
Epoch &    date& $\Delta$R.A.$\times$cos(Dec) &$\Delta$DEC. & JD  &  $\Phi$($P_1$)  &  $\Phi$($P_2$) & $\Phi$($P_{\rm long}$)\\
      &    year-mn-dy-hr-mm& (mas) & (mas) & (day)  &    &   & \\
\hline
 A &2006-06-30-15-26&   2.40&   -0.60&   2453917.143056   &   0.187  & 0.822 &  0.352\\
 B &2006-07-03-15-13&   1.60&   -0.60&   2453920.134028   &   0.300  & 0.934 &  0.354\\
 C &2006-07-07-14-58&   1.40&   -0.40&   2453924.123611   &   0.451  & 0.082 &  0.356\\
 D &2006-07-11-13-42&   1.00&   -0.20&   2453928.070833   &   0.600  & 0.228 &  0.358\\
 E &2006-07-15-14-27&  -0.20&   -0.60&   2453932.102083   &   0.752  & 0.378 &  0.361\\
 F &2006-07-18-14-15&   0.60&    0.40&   2453935.093750   &   0.865  & 0.489 &  0.363\\
 G &2006-07-21-14-03&   0.20&    1.00&   2453938.085417   &   0.978  & 0.600 &  0.364\\
 H &2006-07-24-13-51&   1.40&    1.00&   2453941.077083   &   0.091  & 0.711 &  0.366\\
 I &2006-07-27-13-39&   2.40&   -0.40&   2453944.068750   &   0.203  & 0.822 &  0.368\\
 J &2006-07-30-13-28&   1.80&   -0.60&   2453947.061111   &   0.316  & 0.934 &  0.370\\
\hline                                                                        
 A &2015-07-24-13-34&  -7.45&   -3.10&   2457228.065278   &   0.147  & 0.784 &  0.345\\
 B &2015-07-26-13-27&  -5.55&   -4.80&   2457230.060417   &   0.222  & 0.858 &  0.346\\
 C &2015-07-30-13-11&  -5.70&   -5.05&   2457234.049306   &   0.372  & 0.007 &  0.349\\
 D &2015-08-03-09-25&  -6.70&   -4.50&   2457237.892361   &   0.517  & 0.149 &  0.351\\
 E &2015-08-06-12-43&  -6.95&   -4.35&   2457241.029861   &   0.636  & 0.266 &  0.353\\
 F &2015-08-10-12-28&  -7.15&   -3.90&   2457245.019444   &   0.786  & 0.414 &  0.355\\
 G &2015-08-13-12-15&  -7.15&   -3.85&   2457248.010417   &   0.899  & 0.525 &  0.357\\
 H &2015-08-19-11-52&  -5.70&   -4.90&   2457253.994444   &   0.125  & 0.747 &  0.361\\
 I &2015-08-21-11-44&  -5.30&   -4.85&   2457255.988889   &   0.200  & 0.821 &  0.362\\
 J &2015-08-23-11-36&  -4.85&   -5.50&   2457257.983333   &   0.276  & 0.895 &  0.363\\
\hline
\multicolumn{8}{l}{{\bf{Note}}: Orbital, precessing and long beating phases (turns) are estimated 
with formula $\phi$~=~[($t-t_0$)~mod~$P$]/$P$, by using }\\
\multicolumn{8}{l}{$P_1$~=~26.496 d, $P_2$~=~26.926 d and  $P_{\rm long}$~=~1661 d. The reference 
time t$_0$ is JD 2 443 366.775 \citep{2002ApJ...575..427G}.} 
\end{tabular}
\end{table*}

\newpage
\end{document}